\documentclass[a4paper,11pt]{article}
\usepackage{amssymb}
\usepackage{amsmath}
\usepackage{fullpage}

\makeatletter \@addtoreset{equation}{section} \makeatother

\begin{document}

\begin{titlepage}
 \thispagestyle{empty}
\begin{flushright}
     \hfill{CERN-PH-TH/2010-204}\\
     \hfill{SU-ITP-10/27}
 \end{flushright}

 \vspace{50pt}

 \begin{center}
     { \huge{\bf      {Matrix Norms, BPS Bounds \\  \vspace{5pt}and Marginal Stability in $\mathcal{N}=8$ Supergravity
    }}}

     \vspace{50pt}

     {\Large {Sergio Ferrara$^{a,b,c}$ and Alessio Marrani$^{d}$ }}

     \vspace{30pt}

  {\it ${}^a$ Physics Department, Theory Unit, CERN,\\
     CH -1211, Geneva 23, Switzerland;\\
     \texttt{sergio.ferrara@cern.ch}}

     \vspace{10pt}

    {\it ${}^b$ INFN - Laboratori Nazionali di Frascati,\\
     Via Enrico Fermi 40, I-00044 Frascati, Italy}

     \vspace{10pt}

     {\it ${}^c$  Department of Physics and Astronomy,\\
University of California, Los Angeles, CA 90095-1547,USA}\\

     \vspace{10pt}

   {\it ${}^d$ Stanford Institute for Theoretical Physics,\\
     Stanford University, Stanford, CA 94305-4060,USA;\\
     \texttt{marrani@lnf.infn.it}}

     \vspace{15pt}

     \vspace{80pt}

     {ABSTRACT}

 \vspace{10pt}
 \end{center}
We study the conditions of marginal stability for two-center
extremal black holes in $\mathcal{N}$-extended supergravity in four
dimensions, with particular emphasis on the $\mathcal{N}=8$ case.

This is achieved by exploiting triangle inequalities satisfied by
matrix norms. Using different norms and relative bounds among them,
we establish the existence of marginal stability and split attractor
flows both for BPS and some non-BPS solutions.

Our results are in agreement with previous analysis based on
explicit construction of multi-center solutions.

\end{titlepage}

\section{\label{Intro}Introduction}

In the present investigation we consider BPS bounds for $\mathcal{N}\left(
\geqslant 2\right) $-extended supergravity theories, in connection with the
marginal stability bound of two-center black holes (BHs). Our analysis is
mainly devoted to $\mathcal{N}>2$ theories, since a vast literature and
various results are known for the $\mathcal{N}=2$ case (see \textit{e.g.}
\cite{D-1}-\nocite{D-2,BD-1,KSS-1,DM-1,G-1,WC-1,Gimon-1,CS-1,David-1,GP-1}
\cite{WC-2}; for studies on $\mathcal{N}>2$, see \textit{e.g.} \cite{FGK-1}-
\nocite{Sen,NB,B-1}\cite{B-2}). Within this latter framework, the most
popular application is provided by Calabi-Yau compactifications of (type $II$%
) superstrings. This led to the discovery of the phenomenon of split
attractor flow for multi-center BHs \cite{D-1}, which are stable BPS
solutions, possibly decaying into single center BHs when the scalar flow
cross the wall of \textit{marginal stability }(besides Refs. cited above,
see also \textit{e.g.} \cite{Marginal-Refs}).

In this note we first extend the BPS bound to situations in which the
central charge is an antisymmetric complex matrix $Z_{AB}\left( \phi
,Q\right) $ rather than a complex function. For BPS configurations, as well
as for some non-BPS ones, this is achieved by exploiting Cauchy-Schwarz
triangular inequalities for \textit{matrix norms} of various type (see
\textit{e.g. }\cite{MN}). For instance, in the more familiar case of BPS
bound, the so-called \textit{spectral norm} of $Z_{AB}$ is used.

Interestingly, in some $\mathcal{N}=2$ as well as $\mathcal{N}>2$ theories,
we find double-center non-BPS BH solutions which exhibit a stability region
across a wall of marginal stability. This is ultimately due to the fact that
non-BPS BHs are supported by different charge orbits; in the case of $%
\mathcal{N}=2$ non-BPS solutions with positive quartic $G$-invariant ($%
\mathcal{I}_{4}>0$), similar properties to BPS cases can be found. This is
actually not surprising, because many non-BPS $\mathcal{N}=2$ BH solutions
may become BPS when embedded in higher $\mathcal{N}$ supergravities. In
fact, our analysis, both for BPS and non-BPS cases, agrees with results on
explicit multi-center BPS and non-BPS solutions \cite
{FGK-1,Sen,G-1,Gimon-1,CS-1,GP-1,NB,B-1,B-2}.

In order to use the Cauchy-Schwarz inequality, the crucial point is to
associate the first order ``fake'' superpotential $W$ \cite
{CD,ADOT-1,FO,CDFY-2} to some well-defined matrix norm $\left\| \mathbf{Z}%
\right\| $ of the central charge matrix $\mathbf{Z}$, or of some other
charge matrix. Clearly, when this procedure is possible also for non-BPS
states, the matrix norm under consideration will be different from the
spectral norm $\left\| \mathbf{Z}\right\| _{s}$ which, as mentioned above,
pertains to BPS states.\smallskip

The paper is organized as follows.

In Sec. \ref{Sec.2} we discuss the BPS marginal stability in $\mathcal{N}$%
-extended supergravity by using the spectral norm of $Z_{AB}$. Then, within $%
\mathcal{N}=8$ maximal theory, we derive a manifestly $U$-duality invariant
expression for the marginal stability wall, as well as for the stability
equation fixing the relative distance between the two centers of the
solution in terms of moduli $\phi $ and charges $Q$ (with resulting
non-vanishing overall angular momentum).

In Sec. \ref{Sec.3} we consider several examples in $\mathcal{N}=2$ and $%
\mathcal{N}>2$ supergravity, in which the results derived in Sec. \ref{Sec.2}
hold for non-BPS BHs, as well. In $\mathcal{N}=2$, these include special
K\"{a}hler geometry with $C_{ijk}=0$, as well as the non-BPS states with $%
\mathcal{I}_{4}>0$ in theories with homogeneous symmetric vector multiplets'
scalar manifolds.

Sec. \ref{Sec.4} is instead devoted to the study of the more intriguing case
of non-BPS states with $\mathcal{I}_{4}<0$. Most of the results of our
investigation reproduce the findings of \cite{G-1,GP-1}, namely both the
two-center BH and the two one-center BHs produced by the its decay lie on
the marginal stability wall, and thus no stable region for multi-center
solution exists other than the marginal one. This is related to the fact
that, in these examples, the charge vectors $Q_{1}$ and $Q_{2}$ of the two
centers are \textit{mutually local} (namely, their symplectic product
vanishes: $\left\langle Q_{1},Q_{2}\right\rangle =0$).

A non-BPS $\mathcal{I}_{4}<0$ stable double-center BHs can be found, at
least in $\mathcal{N}=8$ supergravity. This is the case in which the
Pfaffian $Pf\left( \mathbf{Z}\right) $ is real, thus with phase $\varphi
=\pi $, all along the attractor flow. In fact, under this assumption, the
non-BPS ``fake'' superpotential $W_{nBPS}$ can be associated to the
so-called \textit{trace norm} of $\mathbf{Z}$ itself. On the other hand, as
recently shown in \cite{Bena-1}, multi-center non-BPS BHs with constrained
positions of the centers and $\mathcal{I}_{4}<0$ (and therefore non-BPS also
when uplifted to $\mathcal{N}=8$) have been explicitly constructed. It would
be interesting to investigate the occurrence of the split attractor flow in
this framework.

\section{\label{Sec.2}BPS Bounds and Matrix Norms}

We here consider the generalization of the BPS bound as well as of the
Cauchy-Schwarz triangle inequality, which is at the basis of the concept of
\textit{marginal stability}. In order to study this problem, we make a small
prelude on matrix norms (see \textit{e.g.} \cite{MN} for further details).

\subsection{\label{Sec.2.1}Matrix Norms}

Given a complex rectangular $n\times m$ matrix $\mathbf{Z}$, its matrix norm
$\left\| \mathbf{Z}\right\| $ is a consistent generalization of the concept
of vector norm, satisfying by definition the following properties: $\left\|
\mathbf{Z}\right\| \geqslant 0$ ($=0$ \textit{iff} $\mathbf{Z}=0$); $\left\|
\alpha \mathbf{Z}\right\| =\left| \alpha \right| \left\| \mathbf{Z}\right\| $
$\forall \alpha \in \mathbb{C}$; and
\begin{equation}
\left\| \mathbf{Z}_{1}+\mathbf{Z}_{2}\right\| \leqslant \left\| \mathbf{Z}%
_{1}\right\| +\left\| \mathbf{Z}_{2}\right\| .  \label{ti}
\end{equation}

In our treatment, we will be mainly concerned of three types of norms, which
are particular cases of the so-called \textit{Schatten} $p$-norms. Such
matrix norms are defined as the norms of the real vector $\mathbf{\sigma }$
of the \textit{singular values} of a square $n\times n$ matrix $\mathbf{Z}$
(which are nothing but the absolute values of the eigenvalues of $\mathbf{Z}$
itself: $\mathbf{\sigma \equiv }\left\{ \sigma _{i}\right\} _{i=1,..,n}$):
\begin{equation}
\left\| \mathbf{Z}\right\| _{p}\equiv \left( \sum_{i}\sigma _{i}^{p}\right)
^{1/p}.  \label{Schatten-p}
\end{equation}
Namely, we will consider the following norms:

\begin{enumerate}
\item  \textit{Spectral norm}. Starting from the square matrix $\mathbf{Z}$,
one can define the positive semi-definite matrix $\mathbf{ZZ}^{\dag }$,
whose real positive eigenvalues $\lambda _{i}$'s ($i=1,...,m$) are the
squared \textit{singular values} of $\mathbf{Z}$ itself: $\lambda
_{i}=\sigma _{i}^{2}$. The spectral norm $\left\| \mathbf{Z}\right\| _{s}$
of $\mathbf{Z}$ is defined as the \textit{maximum norm} of the vector $%
\mathbf{\sigma }$:
\begin{equation}
\left\| \mathbf{Z}\right\| _{s}\equiv \left\| \mathbf{\sigma }\right\|
_{\infty }\equiv \max \left\{ \sigma _{i}\right\} \equiv \sqrt{\lambda _{h}},
\label{spectral}
\end{equation}
where $\lambda _{h}$ is the highest eigenvalue of the matrix $\mathbf{ZZ}%
^{\dag }$. The spectral norm is formally obtained as the $p\rightarrow
\infty $ limit of the Schatten matrix $p$-norm (\ref{Schatten-p}).

\item  \textit{Frobenius norm}. The Frobenius norm $\left\| \mathbf{Z}%
\right\| _{F}$ of the square matrix $\mathbf{Z}$ is defined as the Euclidean
norm of the vector $\mathbf{\sigma }$ :
\begin{equation}
\left\| \mathbf{Z}\right\| _{F}\equiv \left\| \mathbf{\sigma }\right\|
_{2}\equiv \sqrt{\sum_{i}\lambda _{i}}\equiv \sqrt{Tr\left( \mathbf{ZZ}%
^{\dag }\right) }.
\end{equation}
The Frobenius norm is actually a Schatten matrix $2$-norm. As a particular
case in which the matrix $\mathbf{Z}$ degenerates to complex vector $Z_{I}$,
we will also consider the usual Euclidean norm of $Z_{I}$ ($I=1,...,m$)
itself, defined as
\begin{equation}
\left\| Z_{I}\right\| _{2}\equiv \sqrt{Z_{I}\overline{Z}^{I}}.  \label{Eucl}
\end{equation}

\item  \textit{Trace (or nuclear) norm}. The trace norm $\left\| \mathbf{Z}%
\right\| _{\ast }$ of the square matrix $\mathbf{Z}$ is defined as the $1$%
-norm of the vector $\mathbf{\sigma }$:
\begin{equation}
\left\| \mathbf{Z}\right\| _{\ast }\equiv \left\| \mathbf{\sigma }\right\|
_{1}\equiv \sum_{i}\sqrt{\lambda _{i}}\equiv Tr\left( \sqrt{\mathbf{ZZ}%
^{\dag }}\right) .  \label{trace}
\end{equation}
The trace norm is actually a Schatten matrix $1$-norm.
\end{enumerate}

The crucial property of all these norms is the Cauchy-Schwarz triangle
inequality (\ref{ti}), wich we will exploit in order to study the marginal
stability of double-center BH configurations, in the case in which the
(spatial asymptotical limit of the) relevant matrix norm $\left\| \mathbf{Z}%
\right\| $ is associated to the ADM mass $M_{ADM}$ \cite{ADM} of the BH
solution itself.

The equivalence of the spectral and Frobenius matrix norms is expressed by
the following chain of inequalities:
\begin{equation}
\left\| \mathbf{Z}\right\| _{s}\leqslant \left\| \mathbf{Z}\right\|
_{F}\leqslant \sqrt{rank\left( \mathbf{Z}\right) }\left\| \mathbf{Z}\right\|
_{s}.  \label{te-s-F}
\end{equation}
Let us specify (\ref{te-s-F}) for $\mathbf{Z}$ being the central charge
matrix of $\mathcal{N}=8$, $d=4$ supergravity. In this case, $rank\left(
\mathbf{Z}\right) =8$, and
\begin{equation}
\left\| \mathbf{Z}\right\| _{F}=\sqrt{2\sum_{i=1}^{4}\lambda _{i}}=\sqrt{%
2V_{BH}},
\end{equation}
where $V_{BH}$ is the BH effective potential. Furthermore, due to the
antisymmetry of $\mathbf{Z}$ itself, the Bloch-Messiah-Zumino Theorem \cite
{BMZ} implies the eigenvalues of $\mathbf{Z}$ and $\mathbf{ZZ}^{\dag }$ to
be pairwise; thus, for $\mathbf{Z}$ the chain of inequalities (\ref{te-s-F})
can be made more strict:
\begin{equation}
\left\| \mathbf{Z}\right\| _{s}\leqslant \frac{\left\| \mathbf{Z}\right\|
_{F}}{\sqrt{2}}\leqslant \sqrt{\frac{rank\left( \mathbf{Z}\right) }{2}}%
\left\| \mathbf{Z}\right\| _{s}.  \label{te-s-F-for-Z}
\end{equation}
(\ref{te-s-F-for-Z}) can be rewritten as
\begin{equation}
\sqrt{\lambda _{h}}\leqslant \sqrt{V_{BH}}\leqslant 2\sqrt{\lambda _{h}}%
\Leftrightarrow \lambda _{h}\leqslant V_{BH}\leqslant 4\lambda _{h}.
\label{te-s-F-N=8-d=4}
\end{equation}
This can be extended to the non-BPS case, by noticing that the first order
``fake'' superpotential $W_{nBPS}$ satisfies the bound
\begin{equation}
\left\| \mathbf{Z}\right\| _{s}<W_{nBPS}\leqslant \frac{\left\| \mathbf{Z}%
\right\| _{F}}{\sqrt{2}}\leqslant 2\left\| \mathbf{Z}\right\| _{s},
\label{add-1}
\end{equation}
where the first upper bound on $W_{nBPS}$ is due to Eq. (\ref{V}) further
below. If one further applies (\ref{te-s-F-for-Z}) to the quantity $%
W_{nBPS}\left( \phi ,Q_{1}+Q_{2}\right) $ and uses the triangle inequality
for $\left\| \mathbf{Z}\right\| _{s}$, the following non-BPS inequality is
obtained:
\begin{eqnarray}
W_{nBPS}\left( \phi ,Q_{1}+Q_{2}\right) &\leqslant &2\left\| \mathbf{Z}%
\left( \phi ,Q_{1}+Q_{2}\right) \right\| _{s}\leqslant 2\left[ \left\|
\mathbf{Z}\left( \phi ,Q_{1}\right) \right\| _{s}+\left\| \mathbf{Z}\left(
\phi ,Q_{2}\right) \right\| _{s}\right]  \notag \\
&<&2\left[ W_{nBPS}\left( \phi ,Q_{1}\right) +W_{nBPS}\left( \phi
,Q_{2}\right) \right] .  \label{non-BPS-bound}
\end{eqnarray}
In the spatial asymptotical limit, (\ref{non-BPS-bound}) is an upper limit
for the two-center ADM\ mass in terms of the ADM\ masses of the
single-center constituents. Note that (\ref{non-BPS-bound}) is twice the
marginal stability bound, and in some cases it overestimates the actual
bound. Indeed, for $\mathcal{N}<8$ non-BPS BHs with $\mathcal{I}_{4}>0$ (see
Sec. \ref{Sec.3} further below) the bound satisfied by the corresponding
first order ``fake'' superpotential $W_{\mathcal{I}_{4}>0}$ is a triangle
inequality:
\begin{equation}
W_{\mathcal{I}_{4}>0}\left( \phi ,Q_{1}+Q_{2}\right) \leqslant W_{\mathcal{I}%
_{4}>0}\left( \phi ,Q_{1}\right) +W_{\mathcal{I}_{4}>0}\left( \phi
,Q_{2}\right)
\end{equation}
as in the BPS cases, implying that a stability region for double-center
solutions exists in this case.

It is also interesting to compare (\ref{add-1}) with the chain of
inequalities obtained in \cite{BFK-1}. The lowest bound of (\ref{add-1})
holds for BPS saturation ($W^{2}=\lambda _{h}=\left\| \mathbf{Z}\right\|
_{s}^{2}$), while its highest bound is reached at non-BPS $\mathcal{I}_{4}<0$
attractor points. Thus, the inequality obtained in \cite{BFK-1} is nothing
but the equivalence of the spectral and Frobenius norms of the central
charge matrix $\mathbf{Z}$ of $\mathcal{N}=8$, $d=4$ supergravity.

\subsection{\label{Sec.2.2}BPS Bounds and First Order Flows}

Let us now consider $\mathbf{Z}$ to be the antisymmetric central charge
matrix $Z_{AB}$ ($A,B=1,...,\mathcal{N}$) centrally extending the local
supersymmetry algebra of an $\mathcal{N}$-extended supergravity theory in $d$
space-time dimensions. From the Bloch-Messiah-Zumino Theorem \cite{BMZ}, the
positive semi-definite matrix $\mathbf{ZZ}^{\dag }$ has $\left[ \mathcal{N}/2%
\right] $ independent eigenvalues $\lambda _{i}$ ($i=1,...,\left[ \mathcal{N}%
/2\right] $), and the BPS bound reads (at spatial infinity)
\begin{equation}
M_{ADM}\geqslant \lambda _{h},
\end{equation}
where $M_{ADM}$ denotes the ADM mass of the considered BH state, whereas, as
previously mentioned, $\lambda _{h}\equiv \max \left\{ \lambda _{i}\right\} $%
. If the BPS bound is saturated by $k$ equal highest eigenvalues of $\mathbf{%
Z}$, then the corresponding state is called $\frac{k}{\mathcal{N}}$-BPS. In $%
d=4$ supergravity, if $k>1$ the corresponding BH solution has\footnote{%
In $\mathcal{N}=8$ supergravity for $k=1$ also a ``small'' ($\mathcal{I}%
_{4}=0$) charge orbit exists \cite{FGM,CFMZ1}. This orbit gives both BPS and
non-BPS ``small'' orbits in $\mathcal{N}=2$ theories. No ``small'' non-BPS
orbits exist in $\mathcal{N}=8$.} $\mathcal{I}_{4}=0$ and the near-horizon
space-time geometry is singular (at least in the Einsteinian approximation).
Indeed, it is here worth recalling that the absolute value of the quadratic $%
G$-invariant (if any) $\mathcal{I}_{2}$ or the square root of the absolute
value of the quartic $G$-invariant $\mathcal{I}_{4}$ is the critical,
attractor value of $W^{2}$ of the corresponding flow; thus, through the
Bekenstein-Hawking entropy-area formula \cite{BH}, in the Einstein
supergravity approximation the entropy of the single-center extremal BH
solution reads \cite{KK,FK-Un}
\begin{equation}
S_{BH}=\pi \frac{A_{H}}{4}=\pi \left. W^{2}\right| _{\partial W=0}=\pi
\left. V_{BH}\right| _{\partial V_{BH}=0}=\pi \mathcal{I}\text{,}
\end{equation}
where $A_{H}$ is the area of the BH event horizon, and $\mathcal{I}$ is the $%
G$-invariant ($G$ denoting the $U$-duality group), which does depend on
charges, but not on scalar fields. In the theories under consideration in
the present paper, $\mathcal{I}=\sqrt{\left| \mathcal{I}_{4}\right| }$,
where $\mathcal{I}_{4}$ is the $G$-invariant quartic in charges (as in $%
\mathcal{N}=8$ supergravity), or $\mathcal{I}=\mathcal{I}_{2}$, where $%
\mathcal{I}_{2}$ is the $G$-invariant quadratic in charges (as in $\mathcal{N%
}=2$ \textit{minimally coupled} $\mathbb{CP}^{n}$ models and in $\mathcal{N}%
=3$ supergravity \cite{ADF-U-duality-d=4}).

For extremal BHs, the warp factor of the metric and the scalar flow
associated with the $\frac{k}{\mathcal{N}}$-BPS solution are determined by
the superpotential $W=\sqrt{\lambda _{h}}$, which satisfies first order flow
equations \cite{CD}:
\begin{equation}
\dot{U}=-e^{U}W;~~\dot{\phi}^{\alpha }=-2e^{U}g^{\alpha \beta }\partial
_{\beta }W,
\end{equation}
with the effective BH potential given by
\begin{equation}
V_{BH}=W^{2}+2g^{\alpha \beta }\left( \partial _{\alpha }W\right) \partial
_{\beta }W.  \label{V}
\end{equation}
Note that in $\mathcal{N}=8$, (\ref{V}) can be re-written as a differential
relation between the spectral and Frobenius norms of the central charge
matrix $\mathbf{Z}$:
\begin{equation}
\left\| \mathbf{Z}\right\| _{F}^{2}=2\left\| \mathbf{Z}\right\|
_{s}^{2}+4g^{\alpha \beta }\left( \partial _{\alpha }\left\| \mathbf{Z}%
\right\| _{s}\right) \partial _{\beta }\left\| \mathbf{Z}\right\| _{s}.
\label{V-bis}
\end{equation}

The same relations hold true for non-BPS BHs for all $\mathcal{N}\geqslant 3$
theories and for $\mathcal{N}=2$ models based on symmetric scalar manifolds
(for generalizations beyond symmetric spaces, see \textit{e.g.} \cite{CDFY-2}%
), provided one replaces $W$ with the suitable non-supersymmetric first
order ``fake'' superpotential $W_{nBPS}$ \cite{CD,ADOT-1,FO}. For non-BPS
BHs supported by generic charge configurations with $\mathcal{I}_{4}<0$, the
``fake'' superpotential has a complicated expression (see the first, second
and fourth of Refs. \cite{FO}, and \cite{CDFY-2}). On the other hand, for
all non-BPS BHs with $\mathcal{I}_{4}>0$ the ``fake'' superpotential can be
easily written in terms of a matrix or vector norm of quantities linear in
the charges $Q$. This allows for an analysis of the marginal stability
properties also for such a class of non-BPS constituents and non-BPS
composites.

\subsection{\label{Sec.2.3}BPS Marginal Stability for $\mathcal{N}>2$}

We are now going to apply the triangle inequality (\ref{ti}) of the matrix
norms to the appropriate matrices relevant for the study of extremal BHs in $%
\mathcal{N}$-extended supergravity theories. As mentioned above, for BPS
states, regardless their BPS fraction, the relevant object is the $\mathcal{N%
}\times \mathcal{N}$ complex antisymmetric central charge matrix $\mathbf{Z}%
\equiv Z_{\left[ AB\right] }\left( \phi ,Q\right) $, which is linear in
charges:
\begin{equation}
Z_{AB}\left( \phi ,Q_{1}+Q_{2}\right) =Z_{AB}\left( \phi ,Q_{1}\right)
+Z_{AB}\left( \phi ,Q_{2}\right) .
\end{equation}
Thus, if one assumes the symplectic charge vectors $Q_{1}+Q_{2}$, $Q_{1}$
and $Q_{2}$ to be all BPS, the triangle inequality for the spectral norm $%
\left\| \mathbf{Z}\right\| _{s}$ defined by (\ref{spectral}) yields (in the
spatial asymptotical limit) a bound on the ADM masses, as follows (we omit
the subscript ``$ADM$'' throughout):
\begin{equation}
M\left( \phi _{\infty },Q_{1}+Q_{2}\right) \leqslant M\left( \phi _{\infty
},Q_{1}\right) +M\left( \phi _{\infty },Q_{2}\right) ,  \label{bound-M}
\end{equation}
with $M^{2}=\lambda _{h}$, and ``$\phi _{\infty }$'' denoting the spatially
asymptotical values of scalar fields. The marginal stability condition
corresponds to the saturation of the bound (\ref{bound-M}). Such a
saturation defines the marginal stability wall as the $\left(
Q_{1},Q_{2}\right) $-dependent \textit{locus} in the (spatially
asymptotical) scalar manifold satisfying the equation
\begin{equation}
\sqrt{\lambda _{h}}\left( \phi _{\infty },Q_{1}+Q_{2}\right) =\sqrt{\lambda
_{h}}\left( \phi _{\infty },Q_{1}\right) +\sqrt{\lambda _{h}}\left( \phi
_{\infty },Q_{2}\right) .  \label{ms-line}
\end{equation}

By considering $\mathcal{N}=8$ supergravity, it is worth recalling that the
eigenvalues of $\mathbf{ZZ}^{\dag }$ are solutions of the (square root of
the) \textit{characteristic equation} \cite{DFL-0-brane}
\begin{equation}
\sqrt{\text{det}\left( \mathbf{ZZ}^{\dag }-\lambda \mathbb{I}\right) }%
=\prod_{i=1}^{4}\left( \lambda -\lambda _{i}\right) =\lambda ^{4}+a\lambda
^{3}+b\lambda ^{2}+c\lambda +d=0,
\end{equation}
where the real coefficients $a,b,c,d$, as well as the explicit expressions
of the $\lambda _{i}$'s are given, in terms of $Tr\left( \mathbf{ZZ}^{\dag
}\right) ^{K}$ ($K=1,...,4$), in \cite{DFL-0-brane} (see also the recent
treatment in \cite{CFMZ1}).

The marginal decay of a ``large'' ($\mathcal{I}_{4}>0$) $\frac{1}{8}$-BPS
two-center BH state into two single-center BPS states ($k_{1},k_{2}=1,2,4$)
\begin{equation}
\frac{1}{8}\text{-BPS~\textit{``large''}}\longrightarrow \frac{k_{1}}{8}%
\text{-BPS~}+~\frac{k_{2}}{8}\text{-BPS\label{(1,k1,k2)}}
\end{equation}
can be studied by using (\ref{ms-line}) and the aforementioned expressions
of $\lambda _{i}$'s. Examples of (\ref{(1,k1,k2)}) with $k_{1}$ and/or $%
k_{2}>1$ have been considered \textit{e.g.} in \cite{BD-1,CS-1}.

We note that, since $\sqrt{\left| \mathcal{I}_{4}\left( Q_{1}+Q_{2}\right)
\right| }\neq \sqrt{\left| \mathcal{I}_{4}\left( Q_{1}\right) \right| }+%
\sqrt{\left| \mathcal{I}_{4}\left( Q_{2}\right) \right| }$, the two-center
solution can have less or more entropy than the single-center solution with
the same charge vector $Q_{1}+Q_{2}$. While the BPS single-center BH does
not exist if $\mathcal{I}_{4}\left( Q_{1}+Q_{2}\right) <0$, in the cases
discussed \textit{e.g.} in \cite{BD-1,CS-1} the BPS multi-center solution
has $\mathcal{I}_{4}\left( \sum_{i}Q_{i}\right) \gtrless 0$, but its entropy
is always given by $\sum_{i}\sqrt{\mathcal{I}_{4}\left( Q_{i}\right) }$,
with $\mathcal{I}_{4}\left( Q_{i}\right) \geqslant 0$ $\forall i$.

When at least one of the final two single-center BH states is non-BPS,
namely for cases
\begin{eqnarray}
\frac{1}{8}\text{-BPS~\textit{``large''}} &\longrightarrow &\frac{k_{1}}{8}%
\text{-BPS~}+~\text{nBPS;} \\
\frac{1}{8}\text{-BPS~\textit{``large''}} &\longrightarrow &\text{nBPS~}+~%
\text{nBPS,}
\end{eqnarray}
there is no marginal decay. Indeed, due to the non-saturation of the BPS
bound by one center or both centers, it respectively holds that (at spatial
infinity)
\begin{equation}
\left\| \mathbf{Z}_{1}+\mathbf{Z}_{2}\right\| _{s}\leqslant \left\| \mathbf{Z%
}_{1}\right\| _{s}+\left\| \mathbf{Z}_{2}\right\| _{s}<\left\{
\begin{array}{l}
M_{1}+\left\| \mathbf{Z}_{2}\right\| _{s}; \\
M_{1}+M_{2},
\end{array}
\right.
\end{equation}
where we use the short-hand notation $\mathbf{Z}_{\alpha }\equiv \mathbf{Z}%
\left( Q_{\alpha }\right) $ and $M_{\alpha }\equiv M\left( Q_{\alpha
}\right) $ ($\alpha =1,2$) throughout.

\subsection{\label{Sec.2.4}$\mathcal{N}=8$ BPS Stability Conditions}

Given a two-center BH solution, let us now turn to consider the formula of
the relative distance $\left| \overrightarrow{x_{1}}-\overrightarrow{x_{2}}%
\right| $ of the two single-center BH constituents with \textit{mutually
non-local }charges $\left\langle Q_{1},Q_{2}\right\rangle \neq 0$.

In the $\mathcal{N}=2$ theory (in which $Z_{AB}=Z\epsilon _{AB}$, $A,B=1,2$)
such a distance is \cite{BD-1}
\begin{equation}
\left| \overrightarrow{x_{1}}-\overrightarrow{x_{2}}\right| =\frac{1}{2}%
\frac{\left\langle Q_{1},Q_{2}\right\rangle \left| Z_{1}+Z_{2}\right| }{%
\text{Im}\left( Z_{1}\overline{Z_{2}}\right) },  \label{N=2}
\end{equation}
where $Z_{i}\equiv Z\left( \phi _{\infty },Q_{i}\right) $ ($i=1,2$), and%
\footnote{%
Note that Im$\left( Z_{1}\overline{Z_{2}}\right) =0$ both describes marginal
and \textit{anti-marginal} stability \cite{WC-2}. \textit{Marginal stability
}(at which Re$\left( Z_{1}\overline{Z_{2}}\right) >0$) further requires $%
\left| Z_{1}+Z_{2}\right| ^{2}>\left| Z_{1}\right| ^{2}+\left| Z_{2}\right|
^{2}$. In the other branch $\left| Z_{1}+Z_{2}\right| ^{2}<\left|
Z_{1}\right| ^{2}+\left| Z_{2}\right| ^{2}$, \textit{anti-marginal stability}
(at which Re$\left( Z_{1}\overline{Z_{2}}\right) <0$) corresponds to $\left|
Z_{1}+Z_{2}\right| =\left| \left| Z_{1}\right| -\left| Z_{2}\right| \right| $%
.
\par
All these bounds can be reformulated for $\mathcal{N}>2$ BPS states by
replacing $\left| Z\right| $ with $\sqrt{\lambda _{h}}=\left\| \mathbf{Z}%
\right\| _{s}$.}
\begin{equation}
2\left| \text{Im}\left( Z_{1}\overline{Z_{2}}\right) \right| =\sqrt{4\left|
Z_{1}\right| ^{2}\left| Z_{2}\right| ^{2}-\left( \left| Z_{1}+Z_{2}\right|
^{2}-\left| Z_{1}\right| ^{2}-\left| Z_{2}\right| ^{2}\right) ^{2}}.
\label{N=2--}
\end{equation}

Eq. (\ref{N=2}) implies the stability region for the double-center BH
solution to occur for $\left\langle Q_{1},Q_{2}\right\rangle $Im$\left( Z_{1}%
\overline{Z_{2}}\right) >0$, while it is forbidden for $\left\langle
Q_{1},Q_{2}\right\rangle $Im$\left( Z_{1}\overline{Z_{2}}\right) <0$. Note
that the quantity $\left\langle Q_{1},Q_{2}\right\rangle $Im$\left( Z_{1}%
\overline{Z_{2}}\right) $ is even under the center exchange $%
1\leftrightarrow 2$. The scalar flow is directed from the stability region
towards the instability region, crossing the wall of marginal stability at $%
\left\langle Q_{1},Q_{2}\right\rangle $Im$\left( Z_{1}\overline{Z_{2}}%
\right) =0$. This implies that the stability region is placed \textit{beyond}
the marginal stability wall, and \textit{on the opposite side} of the split
attractor flows.

By using the fundamental identities of $\mathcal{N}=2$ special K\"{a}hler
geometry in presence of two (mutually non-local) symplectic charge vectors $%
Q_{1}$ and $Q_{2}$ (see \textit{e.g.} \cite{D-1,BFM-1,FK-N=8}), one can
compute that at BPS attractor points of the centers $1$ \textit{or} $2$:
\begin{equation}
\mathcal{N}=2:\left\langle Q_{1},Q_{2}\right\rangle =-2\text{Im}\left( Z_{1}%
\overline{Z_{2}}\right) \Rightarrow 2\left\langle Q_{1},Q_{2}\right\rangle
\text{Im}\left( Z_{1}\overline{Z_{2}}\right) =-\left\langle
Q_{1},Q_{2}\right\rangle ^{2}<0.  \label{N=2-1}
\end{equation}
By using (\ref{N=2}) and (\ref{N=2-1}), one obtains $\left| \overrightarrow{%
x_{1}}-\overrightarrow{x_{2}}\right| <0$: this means that, as expected, the
BPS attractor points of the centers $1$ \textit{or} $2$ do not belong to the
stability region of the two-center BH solution. Furthermore, the result (\ref
{N=2-1}) also consistently implies:
\begin{eqnarray}
\text{stability region} &:&\text{~}
\begin{array}{l}
\left\langle Q_{1},Q_{2}\right\rangle \text{Im}\left( Z_{1}\overline{Z_{2}}%
\right) \\
=\left| \left\langle Q_{1},Q_{2}\right\rangle \right| \sqrt{4\left|
Z_{1}\right| ^{2}\left| Z_{2}\right| ^{2}-\left( \left| Z_{1}+Z_{2}\right|
^{2}-\left| Z_{1}\right| ^{2}-\left| Z_{2}\right| ^{2}\right) ^{2}}>0;
\end{array}
\notag \\
&&  \label{stab-region} \\
\text{instability region} &:&\text{~}
\begin{array}{l}
\left\langle Q_{1},Q_{2}\right\rangle \text{Im}\left( Z_{1}\overline{Z_{2}}%
\right) \\
=-\left| \left\langle Q_{1},Q_{2}\right\rangle \right| \sqrt{4\left|
Z_{1}\right| ^{2}\left| Z_{2}\right| ^{2}-\left( \left| Z_{1}+Z_{2}\right|
^{2}-\left| Z_{1}\right| ^{2}-\left| Z_{2}\right| ^{2}\right) ^{2}}<0,
\end{array}
\notag \\
&&  \label{split-flow-region}
\end{eqnarray}
where a particular case of (\ref{split-flow-region}), holding at the
attractor points, is given by (\ref{N=2-1}).

By replacing $\left| Z\right| $ with $\sqrt{\lambda _{h}}$ in (\ref{N=2--}),
the generalization of (\ref{N=2}) to $\mathcal{N}=8$ maximal supergravity
reads
\begin{equation}
\left| \overrightarrow{x_{1}}-\overrightarrow{x_{2}}\right| =\frac{\left|
\left\langle Q_{1},Q_{2}\right\rangle \right| \sqrt{\lambda _{1+2,h}}}{\sqrt{%
4\lambda _{1,h}\lambda _{2,h}-\left( \lambda _{1+2,h}-\lambda _{1,h}-\lambda
_{2,h}\right) ^{2}}},  \label{N=8-gen}
\end{equation}
where $\lambda _{1+2,h}\equiv \lambda _{h}\left( \phi _{\infty
},Q_{1}+Q_{2}\right) $ and $\lambda _{i,h}\equiv \lambda _{h}\left( \phi
_{\infty },Q_{i}\right) $ ($i=1,2$). Note that Eq. (\ref{N=8-gen}) is
manifestly $\mathcal{N}=8$ $U$-duality invariant (written in terms of $%
Tr\left( \mathbf{ZZ}^{\dag }\right) ^{K}$ ($K=1,...,4$)), and it reduces to (%
\ref{N=2}) in the $\mathcal{N}=2$ case. It is here worth remarking that $%
\mathcal{I}_{4}$ of the $\mathcal{N}=8$ theory is a (moduli independent) $%
G=E_{7\left( 7\right) }$-invariant constructed with the (moduli dependent) $%
H=SU(8)$-invariants $Tr\left( \mathbf{ZZ}^{\dag }\right) ^{K}$ ($K=1,2$) and
$Pf\left( \mathbf{Z}\right) $ \cite{CJ,KK,ADF-U-duality-d=4}.

Moreover, a result similar to (\ref{N=2-1}) holds for $\mathcal{N}=8$
supergravity, as well. Indeed, by exploiting the $\mathcal{N}=8$
generalization of the $\mathcal{N}=2$ special geometry identities \cite
{FK-N=8}
\begin{equation}
\left\langle Q_{1},Q_{2}\right\rangle =-\text{Im}\left( Tr\left( \mathbf{Z}%
_{1}\mathbf{Z}_{2}^{\dag }\right) \right) ,  \label{N=8-Ids}
\end{equation}
one can compute that at the $\frac{1}{8}$-BPS attractor points of the
centers $1$ \textit{or} $2$:
\begin{equation}
\mathcal{N}=8:\left| \left\langle Q_{1},Q_{2}\right\rangle \right| =\sqrt{%
4\lambda _{h,1}\lambda _{h,2}-\left( \lambda _{1,h}+\lambda _{2,h}-\lambda
_{1+2,h}\right) ^{2}}.  \label{N=8-gen-1}
\end{equation}
However, note that $\frac{1}{8}$-BPS attractor points of the centers $1$
\textit{or} $2$ do not belong to the stability region of the two-center BH
solution, but instead they are placed, with respect to the stability region,
on the \textit{opposite} side of the marginal stability wall.

It is worth concluding the present Section by remarking that the results (%
\ref{N=2-1}) and (\ref{N=8-gen-1}) are consistent with situations in which
the ADM masses are always on the marginal stability wall (for a given set of
charges, and within a suitable subspace of the scalar manifold, such as
vanishing axions), and then also $\left\langle Q_{1},Q_{2}\right\rangle =0$
(mutually local charges), thus not constraining $\left| \overrightarrow{x_{1}%
}-\overrightarrow{x_{2}}\right| $ in any way (with resulting vanishing
overall angular momentum). For instance, this holds for the limit
(scalarless) case of Reissner-N\"{o}rdstrom double-center BH solutions in $%
\mathcal{N}=2$ pure supergravity. Some other cases are discussed in Sec. \ref
{Sec.4}.

\section{\label{Sec.3}Marginal Stability for Non-BPS $\mathcal{I}_{2}<0$ and
$\mathcal{I}_{4}>0$ Black Holes}

We now consider particular non-BPS double-center BH solutions for which
marginal stability walls can be discussed in full generality.

Let us start with the $\mathcal{N}=2$ theories with $\mathbb{CP}^{n}$ vector
multiplets' scalar manifolds, namely the models in which the $n$ vector
multiplets are \textit{minimally coupled} to the gravity multiplet \cite
{Luciani} (see also \textit{e.g.} \cite{Gnecchi-1}). Such models all have $%
C_{ijk}=0$, and only one type of non-BPS attractors, namely the ones with
vanishing central charge at the horizon ($Z_{H}=0$) and $\mathcal{I}_{2}<0$ (%
$\mathcal{I}_{2}$ denoting the quadratic $G$-invariant of these theories).
The first order ``fake'' superpotential for non-BPS $Z_{H}=0$ is nothing but
the Euclidean norm (\ref{Eucl}) of the complex vector of matter charges $%
Z_{a}\equiv D_{a}Z$ ($a=1,...,n$) in local flat indices \cite{ADOT-1} ($%
\left( z,\overline{z}\right) $ denotes the $\mathcal{N}=2$ - or $\mathcal{N}%
=6$ - , $d=4$ complex scalars throughout):
\begin{equation}
W\left( z,\overline{z};Q\right) =\sqrt{g^{i\overline{j}}Z_{i}\overline{Z}_{%
\overline{j}}}=\sqrt{\sum_{a}\left| Z_{a}\right| ^{2}}=\left\|
D_{a}Z\right\| _{2}.  \label{CP-W-nBPS}
\end{equation}
Thus, due to the linearity of $D_{a}Z$ in the charges $Q$:
\begin{equation}
D_{a}Z\left( z,\overline{z};Q_{1}+Q_{2}\right) =D_{a}Z\left( z,\overline{z}%
;Q_{1}\right) +D_{a}Z\left( z,\overline{z};Q_{2}\right) ,
\end{equation}
the non-BPS $Z_{H}=0$ ``fake'' superpotential given by (\ref{CP-W-nBPS})
satisfies the triangle inequality:
\begin{equation}
W\left( z,\overline{z};Q_{1}+Q_{2}\right) \leqslant W\left( z,\overline{z}%
;Q_{1}\right) +W\left( z,\overline{z};Q_{2}\right) .  \label{PA-1}
\end{equation}
Since the spatial asymptotical limit of $W$ is nothing but the ADM mass
(namely $M\equiv W\left( \phi _{\infty },Q\right) $), it follows that the
saturation of the spatial asymptotical limit of (\ref{PA-1}) yields the
marginal stability condition for the decay
\begin{equation}
\text{\textit{nBPS}}\longrightarrow \text{\textit{nBPS}}+\text{\textit{nBPS}}%
,  \label{nBPS-->nBPS+nBPS}
\end{equation}
with $\mathcal{I}_{2}\left( Q_{1}+Q_{2}\right) <0$, $\mathcal{I}_{2}\left(
Q_{1}\right) <0$ and $\mathcal{I}_{2}\left( Q_{2}\right) <0$.

The same holds true for the unique non-BPS (``large'') charge orbit of $%
\mathcal{N}=3$ supergravity \cite{N=3} (see also \textit{e.g.} \cite
{Gnecchi-1}). This theory also has a quadratic $G$-invariant $\mathcal{I}%
_{2} $, and a first order non-BPS ``fake'' superpotential which is the
Euclidean norm (\ref{Eucl}) of the complex vector of matter charges $Z_{I}$ (%
$I=1,...,n_{V}$, $n_{V}$ denoting the number of vector multiplets) \cite
{ADOT-1}:
\begin{equation}
W\left( z,\overline{z};Q\right) =\left\| Z_{I}\right\| _{2}\equiv \sqrt{Z_{I}%
\overline{Z}^{I}}.  \label{N=3-nBPS}
\end{equation}
Thus, due to the linearity of $Z_{I}$ in the charges $Q$:
\begin{equation}
Z_{I}\left( z,\overline{z};Q_{1}+Q_{2}\right) =Z_{I}\left( z,\overline{z}%
;Q_{1}\right) +Z_{I}\left( z,\overline{z};Q_{2}\right) ,
\end{equation}
the non-BPS $Z_{H}=0$ ``fake'' superpotential given by (\ref{N=3-nBPS})
satisfies the triangle inequality (\ref{PA-1}), whose spatial asymptotical
limit yields an analogue bound for the ADM masses. The saturation of such a
bound is the marginal stability condition for the decay (\ref
{nBPS-->nBPS+nBPS}).

For theories with a quartic $G$-invariant $\mathcal{I}_{4}$, the non-BPS
charge orbit with $\mathcal{I}_{4}>0$ can also be discussed in a fairly
general way. The crucial observation is that this orbit is non-BPS for lower
$\mathcal{N}$'s, but it becomes BPS when the model is embedded in maximal ($%
\mathcal{N}=8$) supergravity. Indeed, it is worth noticing that in $\mathcal{%
N}=8$ supergravity, unlike lower-$\mathcal{N}$ theories, the unique non-BPS
charge orbit is ``large'' with $\mathcal{I}_{4}<0$ \cite{FGM}. Thus, since
the marginal bounds on moduli and charges are insensitive to the value of $%
\mathcal{N}$, the treatment of double-center BH solutions can be performed
(for studies of this issue within a $d=3$ approach, see \cite{NB}).

As an illustrative example, let us consider the $\mathcal{N}=6$ theory,
characterized by the central charge matrix $Z_{AB}$ and a complex singlet
charge $X$. This theory shares the very same bosonic sector with the $%
\mathcal{N}=2$ ``magic'' model based on the degree-$3$ Euclidean Jordan
algebra over the quaternions ($J_{3}^{\mathbb{H}}$), with central charge $%
Z\equiv X$ \cite{ADF,ADF-U-duality-d=4}. After the analysis of \cite{BFGM1},
the $\mathcal{N}=6$ $\frac{1}{6}$-BPS ``large'' orbit becomes the $\mathcal{N%
}=2$ non-BPS $Z_{H}=0$. Thus, the non-BPS $Z_{H}=0$ of the $\mathcal{N}=2$ $%
J_{3}^{\mathbb{H}}$ ``magic'' supergravity has $W=\sqrt{\lambda _{h}}%
=\left\| \mathbf{Z}_{\mathcal{N}=6}\right\| _{s}>\left| X\right| $, and it
satisfies the marginal stability bound because of the triangle inequality on
$\left\| \mathbf{Z}_{\mathcal{N}=6}\right\| _{s}$ itself. In this case, the
formula (\ref{N=8-gen}), clearly with $\lambda _{h}$ denoting the maximal
eigenvalue of the semi-positive definite matrix $\mathbf{Z}_{\mathcal{N}=6}%
\mathbf{Z}_{\mathcal{N}=6}^{\dag }$. On the other hand, the $\mathcal{N}=6$
non-BPS $\mathbf{Z}_{H}\neq 0$ ``large'' orbit corresponds to the $\mathcal{N%
}=2$ ($\frac{1}{2}$-)BPS ``large'' orbit \cite{BFGM1}, with first order
superpotential $\left| Z\right| =\left| X\right| >\sqrt{\lambda _{h}}%
=\left\| \mathbf{Z}_{\mathcal{N}=6}\right\| _{s}$. Thus, due to the
linearity of $X\left( z,\overline{z};Q\right) $ in the charges $Q$, the
triangle inequality (which here is a mere consequence of the Cauchy-Schwarz
inequality on complex numbers)
\begin{equation}
\left| X\left( z,\overline{z};Q_{1}+Q_{2}\right) \right| \leqslant \left|
X\left( z,\overline{z};Q_{1}\right) \right| +\left| X\left( z,\overline{z}%
;Q_{2}\right) \right|
\end{equation}
applies. The relative distance of the two centers $\left| \overrightarrow{%
x_{1}}-\overrightarrow{x_{2}}\right| $ can be computed simply by taking Eq. (%
\ref{N=2}) and replacing $Z$ with $X$.

By exploiting the fact that the complex matter charges' vector $D_{a}Z$ in
local flat indices ($a=1,...,n_{V}$) can be re-arranged in terms of an
antisymmetric complex matrix embedded in the central charge matrix $Z_{AB}$
of $\mathcal{N}=8$ supergravity, one can show the above analysis to hold
true for the non-BPS $\mathcal{I}_{4}>0$ charge orbits of the remaining $%
\mathcal{N}=2$ ``magic'' models (based on $J_{3}^{\mathbb{A}}$, with $%
\mathbb{A}=\mathbb{C},\mathbb{R}$), which are consistent truncation of the
quaternionic model. The ``magic'' octonionic model, based on $J_{3}^{\mathbb{%
O}}$, cannot be obtained through consistent truncation of $\mathcal{N}=8$
theory, but the above analysis can be still shown to hold, since the matter
charges of the $n_{V}=27$ vector multiplets re-arrange (in an $USp\left(
8\right) $-irreducible way) as a skew-traceless $8\times 8$ complex
antisymmetric matrix $Z_{AB}^{0}$, whose skew-trace is the $\mathcal{N}=2$
central charge $Z$.

Therefore, we conclude that the non-BPS $\mathcal{I}_{4}>0$ composites and
constituents may satisfy the marginal stability condition, with a region of
stable double-center BH solutions. Note that this situation is different
from the one discussed in \cite{G-1,GP-1} in which no stable multi-center
configurations were found for non-BPS composites. However, it confirms the
analysis of explicit multi-center non-BPS solutions with $\mathcal{I}_{4}>0$
performed in \cite{NB,B-1,B-2}.

\section{\label{Sec.4}Marginal Stability for Non-BPS $\mathcal{I}_{4}<0$
Black Holes}

In the present Section, we discuss the condition of marginal stability for
non-BPS states with $\mathcal{I}_{4}<0$. In this case, the above reasoning
ascribing the non-BPS lower-$\mathcal{N}$ BH states to BPS orbits in higher-$%
\mathcal{N}$ theories cannot be repeated, because non-BPS BH states with $%
\mathcal{I}_{4}<0$ are all uplifted to non-BPS $\mathcal{I}_{4}<0$ in
maximal supergravity (for investigations within $\mathcal{N}=8$ and $%
\mathcal{N}=4$ theories, see \textit{e.g.} the first and second Refs. of
\cite{NB}).

We actually find that this occurs, in all known examples, in the rather
trivial situation in which the charge vectors $Q_{1}$ and $Q_{2}$ of the two
centers are \textit{mutually local} (\textit{i.e.} $\left\langle
Q_{1},Q_{2}\right\rangle =0$). A non-trivial case is discussed at the end of
the present Section.

It is here worth commenting on the $\mathcal{N}=2$ cases discussed in \cite
{CD}, which are characterized by a``twisted'' (``fake'') central charge. Let
us consider for instance the case discussed, in the ``electric''
configuration $\left( p^{0},q_{1}\right) $ of the so-called $1$-modulus $%
t^{3}$ model, in Sect. 5 therein. In the ($\frac{1}{2}$-)BPS branch ($%
p^{0}q_{1}<0$), the first order superpotential reads
\begin{equation}
W_{BPS}=\left| Z\right| ;~Z=\frac{tq_{1}+p^{0}t^{3}}{\sqrt{-i\left( t-%
\overline{t}\right) ^{3}}},  \label{t^3-BPS}
\end{equation}
while in the non-BPS $Z_{H}\neq 0$ branch ($p^{0}q_{1}>0$) the first order
superpotential reads
\begin{gather}
W_{nBPS}=\left| Z_{twist}\right| ;~Z_{twist}=\frac{tq_{1}+p^{0}t^{2}%
\overline{t}}{\sqrt{-i\left( t-\overline{t}\right) ^{3}}}=t\frac{\left(
q_{1}+p^{0}\left| t\right| ^{2}\right) }{\sqrt{-i\left( t-\overline{t}%
\right) ^{3}}}; \\
\Downarrow  \notag \\
W_{nBPS}=\pm \left| t\right| \frac{\left( q_{1}+p^{0}\left| t\right|
^{2}\right) }{\sqrt{-i\left( t-\overline{t}\right) ^{3}}}\text{~for~}%
p^{0},q_{1}\gtrless 0.  \label{t^3-nBPS}
\end{gather}
Thus, $W_{nBPS}$ given by (\ref{t^3-nBPS}) is linear in charges, whereas $%
W_{BPS}$ given by (\ref{t^3-BPS}) is not:
\begin{eqnarray}
\left| Z_{twist}\left( Q_{1}+Q_{2}\right) \right| &=&\left| Z_{twist}\left(
Q_{1}\right) \right| +\left| Z_{twist}\left( Q_{2}\right) \right| ; \\
\left| Z\left( Q_{1}+Q_{2}\right) \right| &\leqslant &\left| Z\left(
Q_{1}\right) \right| +\left| Z\left( Q_{2}\right) \right| .
\end{eqnarray}
Thus, the twist $t^{3}\rightarrow t^{2}\overline{t}$ determining $%
Z\rightarrow Z_{twist}$ makes the stability region for the two-center
non-BPS configuration empty. The multi-center solutions discussed in \cite
{G-1,GP-1} are of this kind. Note that $\left( p^{0},q_{1}\right) $ is a
closed subspace with respect to charge addition, as in general ``electric''
and ``magnetic'' configurations (discussed further below) are, as well.

In the particular $\mathcal{N}=2$ cases discussed in \cite{CD}, it is
observed that $W_{\mathcal{I}_{4}<0}\left( z,\overline{z};Q\right) $, also
in presence of non-vanishing axions, is \textit{linear} in charges (we omit
the subscript ``$\mathcal{I}_{4}<0$'' throughout):
\begin{equation}
W\left( z,\overline{z};Q_{1}+Q_{2}\right) =W\left( z,\overline{z}%
;Q_{1}\right) +W\left( z,\overline{z};Q_{2}\right) .  \label{W-linear-in-Q}
\end{equation}
The property (\ref{W-linear-in-Q}) has an obvious consequence, namely that
non-BPS double-center configurations always occur at the marginal stability
wall in the moduli space, since the spatial asymptotical limit of (\ref
{W-linear-in-Q}) reads
\begin{equation}
M_{1+2}\left( z_{\infty },\overline{z}_{\infty };Q_{1}+Q_{2}\right)
=M_{1}\left( z_{\infty },\overline{z}_{\infty };Q_{1}\right) +M_{2}\left(
z_{\infty },\overline{z}_{\infty };Q_{2}\right) .
\end{equation}
Therefore, for these charge configurations the non-BPS BH bound states are
never stable but rather only marginally stable, thus producing two
single-center BH solutions with mutually local charges ($\left\langle
Q_{1},Q_{2}\right\rangle =0$) and no constraints on the relative distance $%
\left| \overrightarrow{x_{1}}-\overrightarrow{x_{2}}\right| $ between the
two centers (and therefore vanishing overall angular momentum). A further
example is provided by Eq. (4.1) of the first Ref. of \cite{FO}.

A particular subset of such configurations are the ``electric'' ($%
p^{0},q_{i} $) and ``magnetic'' ($q_{0},p^{i}$) ones, which may be
axion-free. By plugging them into the explicit general expression of $Z$
computed in \cite{CFM1}, one finds that such configurations support a real
or purely imaginary central charge: $Z=\pm \overline{Z}$. As a consequence,
both BPS and non-BPS constituents do not form a stable composite, and the
moduli are always on the marginal stability wall. Notice that the situation
is different for the $\left( p^{0},q_{0}\right) $ charge configuration
(corresponding to the presence of only $D0$ and $D6$ branes \cite{CS-1}).
This configuration may (but does not necessarily) support axion-free
solutions but, as already evident in the $1$-modulus $t^{3}$ model (see
\textit{e.g.} Eq. (3.5) of the first Ref. of \cite{FO}), $W_{\mathcal{I}%
_{4}<0}$ is not linear in charges nor the absolute value of a complex
quantity linear in charges.

Thus, apart from the $\left( p^{0},q_{0}\right) $ case, it seems that many
known simple non-BPS $\mathcal{I}_{4}<0$ configurations are exactly
marginal. This situation agrees with the conclusions of the analysis of \cite
{G-1,GP-1}.

By using norm inequalities, the only non-BPS $\mathcal{I}_{4}<0$
configurations which may exhibit a stability region for double-center BH
solutions (and a corresponding wall of marginal stability for their decay
into two single-center BH solutions) seem to be the ones which can be
uplifted to a very particular non-BPS configurations of $\mathcal{N}=8$
supergravity, namely one with constant phase. In such a case, one of the
duality ($SU\left( 8\right) $-) invariants of the theory, namely the
Pfaffian $Pf\left( \mathbf{Z}\right) $ of the central charge matrix $\mathbf{%
Z}$, is constrained to be real all along the corresponding scalar flow; this
corresponds to the phase $\varphi $ of $\mathbf{Z}$ to be set to its non-BPS
critical value $\varphi _{H}=\pi $ all along the flow. For this
configurations, the first order non-BPS ``fake'' superpotential can be
computed to be nothing but (one quarter of) the trace norm (\ref{trace}) of $%
\mathbf{Z}$ itself \cite{ADOT-1} (see also the second and third Refs. of
\cite{FO}):
\begin{equation}
W_{\mathcal{I}_{4}<0,\varphi =\pi }\left( \phi ,Q\right) =\frac{1}{2}%
\sum_{i=1}^{4}\sqrt{\lambda _{i}}=\frac{1}{4}Tr\left( \sqrt{\mathbf{ZZ}%
^{\dag }}\right) =\frac{1}{4}\left\| \mathbf{Z}\right\| _{\ast }.
\end{equation}
Consequently, $W_{\mathcal{I}_{4}<0,\varphi =\pi }$ satisfies the triangle
inequality
\begin{equation}
W_{\mathcal{I}_{4}<0,\varphi =\pi }\left( \phi ,Q_{1}+Q_{2}\right) \leqslant
W_{\mathcal{I}_{4}<0,\varphi =\pi }\left( \phi ,Q_{1}\right) +W_{\mathcal{I}%
_{4}<0,\varphi =\pi }\left( \phi ,Q_{2}\right) ,  \label{DE-bound}
\end{equation}
provided that (recall (\ref{N=8-Ids}))
\begin{equation}
\left\langle Q_{1},Q_{2}\right\rangle =-\text{Im}\left( Tr\left( \mathbf{Z}%
_{1}\mathbf{Z}_{2}^{\dag }\right) \right) \neq 0,  \label{DE-cond-1}
\end{equation}
and that $Pf\left( \mathbf{Z}_{1}+\mathbf{Z}_{2}\right) $, $Pf\left( \mathbf{%
Z}_{1}\right) $ and $Pf\left( \mathbf{Z}_{2}\right) $ are all real; this
latter condition can equivalently be recast as
\begin{equation}
\varphi \left( \phi ,Q_{1}+Q_{2}\right) =\varphi \left( \phi ,Q_{1}\right)
=\varphi \left( \phi ,Q_{2}\right) =\pi ,  \label{DE-cond-2}
\end{equation}
all along the attractor flow. The marginal stability condition would
correspond to the saturation of the bound (\ref{DE-bound}), within the
conditions (\ref{DE-cond-1}) and (\ref{DE-cond-2}).

By performing the supersymmetry reduction $\mathcal{N}=8\rightarrow \mathcal{%
N}=2$ and using the $\mathcal{N}=2$ formalism introduced in the first Ref.
of \cite{FO} and in \cite{CDFY-2}, the constancy of the phase $\varphi $
along the non-BPS $\mathcal{I}_{4}<0$ attractor flow corresponds to the
vanishing of the $H$-invariant $i_{3}$ (and to $i_{4}<0$). Thus, the $%
\mathcal{N}=2$ analogues of conditions (\ref{DE-cond-1}) and (\ref{DE-cond-2}%
) respectively read as follows (for the equality in the l.h.s. of (\ref
{DE-cond-N=2-1}), see \textit{e.g.} \cite{D-1,BFM-1}):
\begin{eqnarray}
\left\langle Q_{1},Q_{2}\right\rangle &=&2\text{Im}\left[ -Z_{1}\overline{%
Z_{2}}+g^{i\overline{j}}\left( D_{i}Z_{1}\right) \overline{D}_{\overline{j}}%
\overline{Z_{2}}\right] \neq 0;  \label{DE-cond-N=2-1} \\
i_{3}\left( z,\overline{z};Q_{1}+Q_{2}\right) &=&i_{3}\left( z,\overline{z}%
;Q_{1}\right) =i_{3}\left( z,\overline{z};Q_{2}\right) =0,~i_{4}\left( z,%
\overline{z};Q_{1}+Q_{2}\right) <0.  \label{DE-cond-N=2-2}
\end{eqnarray}
The moduli dependence of (\ref{DE-cond-2}) and (\ref{DE-cond-N=2-2}) yields
a co-dimension three subspace of scalar manifold. Thus, in the $\mathcal{N}%
=8\rightarrow \mathcal{N}=2$ supersymmetry reduction, if the three real
conditions entailed by (\ref{DE-cond-N=2-2}) are all independent, they admit
consistent solutions in presence of mutually non-local charges $\left\langle
Q_{1},Q_{2}\right\rangle \neq 0$ only with at least two (complex) scalar
fields.

\section{\label{Conclusion}Concluding Remarks}

In the present investigation, we have analyzed the marginal stability bound
for BPS extremal (two-center composite) BHs in $\mathcal{N}>2$ supergravity,
as well as whether this bound can be extended to non-BPS configurations.

By denoting the central charge matrix with $\mathbf{Z}$, for BPS BHs we
found that the Cauchy-Schwarz triangle inequality applies to the ADM mass $%
M=\lim_{r\rightarrow \infty }\sqrt{\lambda _{h}}=\lim_{r\rightarrow \infty
}\left\| \mathbf{Z}\right\| _{s}$, where $\lambda _{h}$ is the highest
eigenvalue of the semi-positive definite matrix $\mathbf{ZZ}^{\dag }$, and $%
\left\| \cdot \right\| _{s}$ stands for the \textit{spectral} matrix norm.
This generalization of the marginal stability bound uses the property of
matrix norm as well as the linearity of $\mathbf{Z}$ in charges $Q$:
\begin{equation}
\left\| \mathbf{Z}\left( \phi ,Q_{1}+Q_{2}\right) \right\| _{s}=\left\|
\mathbf{Z}\left( \phi ,Q_{1}\right) +\mathbf{Z}\left( \phi ,Q_{2}\right)
\right\| _{s}\leqslant \left\| \mathbf{Z}\left( \phi ,Q_{1}\right) \right\|
_{s}+\left\| \mathbf{Z}\left( \phi ,Q_{2}\right) \right\| _{s}.
\end{equation}

Furthermore, we found that all non-BPS BHs of the $\mathcal{N}=2$ \textit{%
minimal coupling} $\mathbb{CP}^{n}$ sequence (characterized by $C_{ijk}=0$)
and of $\mathcal{N}=3$ supergravity, satisfy a marginal stability bound
identical to the one of their BPS counterparts. These theories share the
properties that they cannot be uplifted to $d=5$ space-time dimensions, they
have a $G$-invariant $\mathcal{I}_{2}$ which is quadratic in charges, which
defines the supersymmetry preserving features of the charge orbits as
follows:
\begin{equation}
\text{BPS}:\mathcal{I}_{2}\geqslant 0;~\text{nBPS}:\mathcal{I}_{2}<0.
\end{equation}

For theories with a $G$-invariant $\mathcal{I}_{4}$ quartic in charges and $%
\mathcal{N}<8$, two types of ``large'' attractor non-BPS solutions exist,
depending on whether $\mathcal{I}_{4}\gtrless 0$.

For $\mathcal{I}_{4}>0$ non-BPS BHs, the marginal stability bound as for the
BPS BHs applies. An obvious example is provided by the $\mathcal{N}=6$
theory, which shares the same bosonic sector of the $\mathcal{N}=2$
``magic'' quaternionic ($J_{3}^{\mathbb{H}}$-based) supergravity, but with
the role of BPS and non-BPS (both with $\mathcal{I}_{4}>0$) interchanged
\cite{BFGM1}. This example actually extends to the $\mathcal{I}_{4}>0$
non-BPS BHs of all $\mathcal{N}\geqslant 2$-extended supergravities with
symmetric (vector multiplets') scalar manifolds. The $\mathcal{N}=5$ case is
particularly simple, because such a theory has only two orbits, both BPS:
one ``large'' ($\frac{1}{5}$-BPS) and one ``small'' ($\frac{2}{5}$-BPS). At
least for ``magic'' $\mathcal{N}=2$ models (with the exclusion of the
octonionic one), this result for $\mathcal{I}_{4}>0$ non-BPS BHs is not
surprising, because such theories can be seen as sub-theories of the maximal
$\mathcal{N}=8$ supergravity, in which in fact the constraint $\mathcal{I}%
_{4}>0$ defines a unique ($\frac{1}{8}$-BPS) orbit \cite{FGM}.

For $\mathcal{I}_{4}<0$ non-BPS BHs, we found that most examples
(characterized by particular charge configurations and moduli dependence)
saturate the marginal stability bound, and thus they cannot admit stable
double-center composite solutions. It would be interesting to determine
under which circumstances, for a generic charge configuration belonging to
the non-BPS ``large'' orbit $\frac{E_{7\left( 7\right) }}{E_{6\left(
6\right) }}$, the $\mathcal{N}=8$ non-BPS first order ``fake''
superpotential, which in the asymptotical spatial limit yields the ADM mass,
satisfies the marginal stability bound. It should be recalled that the
\textit{Ansatz} of flat $d=3$ spatial slices of the BH geometry, made in
\cite{G-1,GP-1}, has been removed in \cite{Bena-1}, in which a general
solution for non-BPS multi-center BHs, with constrained centers and
non-vanishing overall angular momentum, has been explicitly obtained.

For stable configurations with a wall of marginal stability, the split
attractor flow will occur not only for BPS cases, but also for non-BPS cases
for which a stability region in the moduli space exists. In this paper we
have shown that, at least in the supergravity approximation, this is not
limited to BPS solutions, but it extends to a broad class of non-BPS
solutions.

Finally, it would be interesting to investigate, in the case of $\mathcal{N}%
\geqslant 2$ non-BPS and also $\mathcal{N}>2$ BPS configurations, the fate
of the \textit{``moduli spaces''} \cite{Ferrara-Marrani-2} of scalar flows
across the split occurring at the marginal stability wall, which may be thus
reduce or remove the ``flat directions'' spanning the corresponding \textit{%
``moduli space''}. For $\mathcal{N}=8$ $\frac{1}{8}$-BPS (``large'')
attractor flow, the ``flat directions'' have an $\mathcal{N}=2$
interpretation in terms of hypermultiplets' scalar degrees of freedom \cite
{ADF-U-duality-d=4,Ferrara-Marrani-1,Ferrara-Marrani-2}. Therefore, the
double-center solutions removing the ``flat directions'' would be genuine $%
\mathcal{N}=8$ solutions with no $\mathcal{N}=2$ interpretation (concerning
this, see \cite{NB,B-2}).

\section*{Acknowledgments}

We would like to thank V. S. Varadarajan for enlightening discussions.

The work of S. F. is supported by the ERC Advanced Grant no. 226455, \textit{%
``Supersymmetry, Quantum Gravity and Gauge Fields''} (\textit{SUPERFIELDS})
and in part by DOE Grant DE-FG03-91ER40662.

The work of A. M. has been supported by an INFN visiting Theoretical
Fellowship at SITP, Stanford University, CA, USA.

\end{document}